\documentclass[%
reprint,
superscriptaddress,
 amsmath,amssymb,
 aps,
]{revtex4-2}
\usepackage{algorithm}
\usepackage{algpseudocode}
\usepackage{svg}
\usepackage{url}
\usepackage{graphicx}
\usepackage{bm}
\usepackage[page]{appendix} 
\usepackage{lipsum}
\usepackage{hyperref}

\newcommand{\av}[1]{\langle{#1}\rangle{}}
\newcommand{\beq}{\begin{equation}}
\newcommand{\eeq}{\end{equation}}
\newcommand{\beqs}{\begin{equation*}}
\newcommand{\eeqs}{\end{equation*}}
\newcommand{\beqn}{\begin{eqnarray}}
\newcommand{\eeqn}{\end{eqnarray}}
\newcommand{\beqns}{\begin{eqnarray*}}
\newcommand{\eeqns}{\end{eqnarray*}}

\newcommand{\lastequal}{Corresponding authors. These authors contributed equally.}

\newcommand{\EQ}[1]{Eq.~\eqref{#1}}

\newcommand{\ttilde}[1]{\Tilde{\Tilde{#1}}}

\begin{document}

\title{How host mobility patterns shape antigenic escape during viral-immune co-evolution}
\author{Natalie Blot}
\altaffiliation{These authors contributed equally to this work.}
\affiliation{Gulliver Lab UMR CNRS 7083, ESPCI Paris, Universit\'{e} PSL, 75005 Paris, France}
\author{Caelan Brooks}
\altaffiliation{These authors contributed equally to this work.}
\affiliation{Department of Physics, Harvard University, Cambridge, Massachusetts 02138, USA}
\author{Daniel W. Swartz}
\altaffiliation{These authors contributed equally to this work.}
\affiliation{Department of Physics, Massachusetts Institute of Technology, Cambridge, Massachusetts 02139, USA}
\author{Eslam Abdelaleem}
\affiliation{Department of Physics, Emory University, Atlanta, Georgia 30322, USA}
\author{Martin Garic}
\affiliation{Sorbonne Universit\'{e}, CNRS, Institut de Biologie Paris-Seine (IBPS), Laboratoire Jean-Perrin (LJP), F-75005, Paris, France}
\author{Andrea Iglesias-Ramas}
\affiliation{Institut Curie, PSL Research University, Sorbonne Université, CNRS UMR 168, Laboratoire Physique des Cellules et Cancer, 75005 Paris, France.}
\author{Michael Pasek}
\affiliation{Department of Physics, Emory University, Atlanta, Georgia 30322, USA}
\affiliation{Initiative in Theory and Modeling of Living Systems, Emory University, Atlanta, United States}
\author{Thierry Mora}
\altaffiliation{\lastequal}
\author{Aleksandra M. Walczak}
\altaffiliation{\lastequal}
\affiliation{Laboratoire de physique de l’École normale supérieure, CNRS, PSL University, Sorbonne Université, and Université Paris Cité, 75005 Paris, France}

\begin{abstract}
Viruses like influenza have long coevolved with host immune systems, gradually shaping the evolutionary trajectory of these pathogens. Host immune systems develop immunity against circulating strains, which in turn avoid extinction by exploiting antigenic escape mutations that render new strains immune from existing antibodies in the host population. Infected hosts are also mobile, which can spread the virus to regions without developed host immunity, offering additional reservoirs for viral growth. While the effects of migration on long term stability have been investigated, we know little about how antigenic escape coupled with migration changes the survival and spread of emerging viruses. By considering the two processes on equal footing, we show that on short timescales an intermediate host mobility rate increases the survival probability of the virus through antigenic escape. We show that more strongly connected migratory networks decrease the survival probability of the virus. Using data from high traffic airports we argue that current human migration rates are beneficial for viral survival. 
\end{abstract}

\maketitle

\section{Introduction}

Viruses and their hosts have coevolved since the earliest form of cellular life \cite{kajan2020virus}. 
In humans and other jawed vertebrates, infected hosts produce a specific adaptive immune response through the activation and proliferation of specific B and T cells that prevent viral spreading and kill infected cells.
Once the infection has been cleared, hosts retain an immune memory in the form of memory B and T cells, enabling rapid response in the case of reinfection~\cite{sompayrac2022immune,ahmed1996immunological}. This long-lasting immune protection places strong selective pressures on circulating pathogenic strains, driving the coevolution of viral surface proteins in the face of ever adapting immune protections~\cite{McCoy2015, Phillips2021a, Mazzolini2023, Chardes2022, Thyagarajan2014}. 
 
The evolution of the influenza virus is an example of the coevolutionary feedback experienced by viruses and the population of its host immune systems~\cite{Thyagarajan2014, morris2018predictive, sakai2022history, Luksza2014}.
Globally circulating influenza strains compete for hosts which produces a selective force acting on the genetic variation within the influenza population~\cite{Bedford2010,Strelkowa2012}.   
Together, selection and viral mutations occurring at high rates fuel the rapid emergence of new antigenic variants that in turn lead to the evolution of host immune systems that acquire protection~\cite{Bloom2010, Thyagarajan2014, Doud2018, Phillips2021a}. Recent in-lab expression of historical 20th century influenza strains shows that the virus did undergo antigenic escape~\cite{Lee2018, Doud2018}. 

The influenza virus has been stably evolving in humans for a long time~\cite{sakai2022history}, finding itself now in a coevolutionary steady state~\cite{petrova2018evolution}. While selection pressures acting on influenza B strains has led to the coexistence of two stable lineages, Victoria and Yamagata, the evolution of influenza A/H3N2 has resulted in one stable lineage since the late 1960s. Dimensionality reduction of strains within the influenza A/H3N2~\cite{Smith2004, Bedford2012} shows that the antigenic escape  is well described in terms of a one dimensional  traveling antigenic wave~\cite{sasaki1994evolution, Bedford2012, Yan2018, Marchi2019, Marchi2021,chardes2023evolutionary}.

Antigenic wave descriptions of coevolution assume a single well-mixed host population, ignoring spatial effects. Large scale  spatial models of epidemiology have been instrumental in guiding policies and explaining epidemic dynamics~\cite{Balcan2009, Luca2018, Pullano2021,Poletto2013}, but usually
do not include antigenic escape.
Host migration is known to be an important factor in influenza spreading \cite{Bedford2010, Bedford2015, Wen2016, Hadfield2018, Tsui2023}. 
The effect of fragmenting a population into many sub-populations is known to strongly affect the persistence of species in different ecological scenarios~\cite{keeling2000metapopulation,fox2017population}, and a similar fragmentation of potential hosts can change epidemiological features of a disease, such as the epidemic threshold \cite{Pastor-Satorras2015}.
However, few models capture all three features---epidemiological dynamics, antigenic escape, and the spatial structure of populations---explicitly. Refs.~\cite{Boni2006, kumata2022antigenic} treat epidemiological dynamics and antigenic escape together in a well-mixed population, while Ref.~\cite{sasaki2022antigenic} explores the evolutionary impact of punctuated antigenic escape. Others  treat population structure and epidemic dynamics~\cite{Jesse_Heesterbeek_2011, Pastor-Satorras2015} or networks structure and evolutionary impact without a host-disease model~\cite{mcmanus2021evolution,lamarins2024eco}. Ecologists have studied ``eco-evo coupling'' with great interest in the past few decades as mounting evidence shows the importance of such phenomena, but often in well-mixed populations and at steady state. 
Lastly, spatial models of stochastic evolution have coupled small population sizes with spatial structure, allowing individuals to migrate between spatially distinct patches which we call demes, and explored the role of migration on population survival~\cite{Houchmandzadeh2011,Slatkin1981, Marrec2021, Kuo2024}, but these models have not been applied to co-evolutionary settings.

Here we focus on antigenic escape in a stochastic co-evolutionary model to study the effects of host migration on the survival of a new viral strain following its outbreak. We ask how the interplay between host spatial migration and evolution within an abstract antigenic space influences the survival probability of the virus (Fig.~\ref{fig:model_schematic}).

\section{Model}

We describe viral-immune co-evolution in terms of a stochastic model of antigenic drift coupled to epidemiological dynamics \cite{Chardes2023} in a population of identical hosts structured in demes.  We focus on mutations that are neutral besides their interaction with the immune system, and do not affect life-history traits like transmission or recovery rates.   Nevertheless these mutations can reduce the binding affinity between antibodies and antigens, effectively allowing the evolved strains to escape host immunity. 

\begin{figure*}
    \centering
    \includegraphics[width=\textwidth]{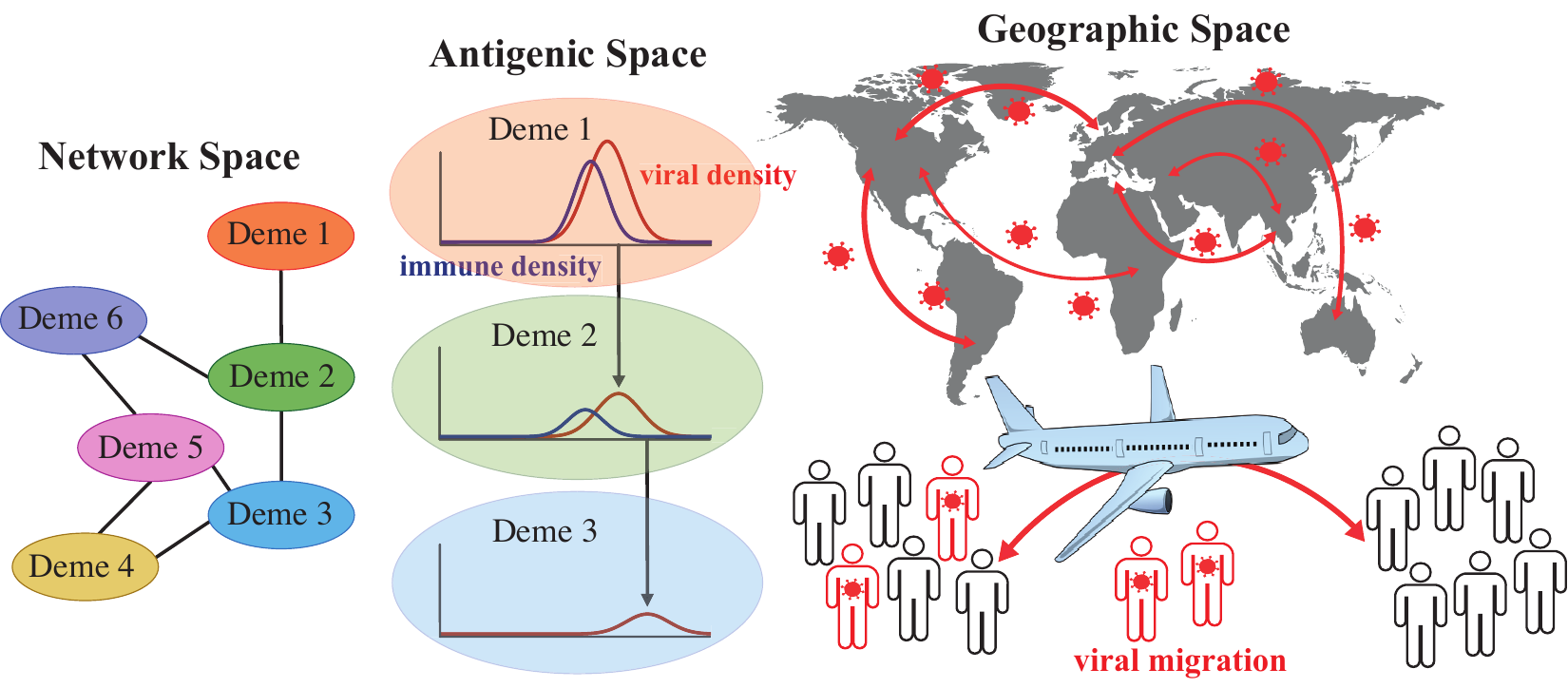}
    \caption{\textbf{Schematic illustrating the different spaces encompassed by the model.} \textbf{(a)} Network space: demes of constant population sizes communicate through host migration. \textbf{(b)} Antigenic space: within each deme, dynamics follow the model described in Eqs. \eqref{viral_diff_eqn_please_work} and \eqref{immune diff eqn}. \textbf{(c)} Geographical space: the connectivity of demes within the network is facilitated by the movement of infected people in geographic space. In section \ref{real data} we connect the spaces in (a) and (b) to the geographic space described in (c).}
    \label{fig:model_schematic}
\end{figure*}

Individuals are well-mixed within each deme, and demes are coupled by migration between them (Fig.~\ref{fig:model_schematic}~A). We describe the interactions between viruses and hosts with a susceptible-infected-recovered (SIR) model. To effectively describe the space of all possible antigenic strains, we introduce a one-dimensional antigenic space~\cite{Segel1989} and  label the viral strains and host immunity by a continuous antigenic coordinate $x$. While antigenic space is generally of higher dimension, canalized evolution makes it effectively one-dimensional~\cite{Bedford_Rambaut_Pascual_2012, Smith2004} if we ignore rare splitting events \cite{Marchi2019, Marchi2021}.
The density of hosts infected by strain $x$ at a given time $t$ in deme $i$ is $n_i(x,t)$, and the fraction of hosts in deme $i$ who are susceptible to $x$ is denoted by $S_i(x)$. Upon introduction into a deme that is entirely susceptible to strain $x$ ($S_i(x,t)=1$), the virus will grow exponentially with a characteristic transmission rate $\beta$. An infected host will mount an adaptive immune response and clear the infection with a recovery rate $\gamma$ or die with rate $\alpha$, such that $\partial n_i(x,t) / \partial t = (\beta S_i(x,t) - \gamma - \alpha) n_i(x,t)$.  For influenza infections in otherwise healthy human hosts recovery is the most common outcome so we set $\alpha=0$.  
 
 A viral strain $x$ accumulates many antigenic mutations that are small in magnitude, such that it effectively diffuses through the antigenic space with a diffusion coefficient $D = (\delta x)^2/2\delta t$, where $\delta x$ is the typical effect size of a single mutation occurring on a typical timescale $\delta t$. The viral population density in antigenic space changes due to fitness, migration and mutations:
\begin{equation} \label{viral_diff_eqn_please_work}
\begin{split}
    \frac{\partial n_i(x,t)}{\partial t} &= ( \beta S_i(x,t) - \gamma - \alpha) n_i \\
    &+ \sum_{j \neq i} \left[ K_{ij} n_j - K_{ji} n_i \right] \\
    &+ D \frac{\partial^2 n_i}{\partial x^2} + \sigma\sqrt{n_i}\xi_i(t), 
\end{split}
\end{equation}
where  $K_{ij}$ is the migration rate from deme $j$ to deme $i$, and $\xi_i(t)$ is a Gaussian white noise of unit amplitude. Unless otherwise specified, $K_{ij}$ is symmetric so that all demes have equal size at steady state. The parameter $\sigma$ corresponds to the standard deviation of the reproductive number (see Appendix).

 In response to the viral strains, hosts evolve a population-level immune density in each deme $i$, $h_i(x,t)$. For numerical simplicity
 we assume that each deme has a constant number of hosts, $N_h$.
 Each  host has $M$ immune protections and a host infected by an unrecognized strain $x$  randomly replaces an existing protection with immunity to $x$, leading to the following dynamics:
\begin{equation}  \label{immune diff eqn}
    \partial_t h_i(x, t)=\frac{1}{M N_h}[n_i(x, t)-N_i(t) h_i(x, t)],
\end{equation}
    where $N_i(t) = \int n_i(x,t) dx$ is the total number of infected hosts in deme $i$.  We assume immunity decays exponentially with antigenic distance, with characteristic length $r_0$. The population-level immune coverage is defined as:
\begin{equation}
    c_i(x, t)=\int  h_i(y, t) e^{-\frac{|x-y|}{r_0}} dy.
    \label{ceqn}
\end{equation}
The probability for a potential host to be susceptible to $x$ is then given by the probability that none of their $M$ memories protect them from strain $x$,
\begin{equation}
    S_i(x, t)=(1-c_i(x, t))^M.
    \label{Seqn}
\end{equation}

This model was previously studied at steady state in the case of a single deme \cite{chardes2023evolutionary}. It was shown to converge to a stationary solution in a frame moving with constant speed. The speed, size, and width of the wave depends on the cross-reactivity range of the immune protection. Two tractable regimes emerge when two time scales of the problem are well separated:
the typical escape time of a mutating virus, and the basic doubling time of the virus in absence of immunity, $\kappa=r_0^2 (\beta - \alpha - \gamma)/D$. For relatively small ratios of these timescales ($\kappa \ll 10^3$), the coevolutionary dynamics is  described by a  Fisher-Kolmogorov-Petrovsky-Piskunov (FKPP) wave, and for large ratios ($\kappa \gg 10^3$) it is described by a linear-fitness wave. Our simulations were done close to the cross-over regime, with parameter choices resulting in $\kappa \sim 10^3$.

In this paper we want to study the outbreak of a new strain to which the population is entirely susceptible. To do so, we initiate all the simulations with no immune coverage, $h_i(x,t=0)=0$, and with a small number of infected hosts appearing in a single deme, $n_1(x,t=0)>0$, and $n_{i\neq 1}(x,t=0)=0$.

\section{Results}


\subsection{One Deme}
\begin{figure*}
    \centering
    \includegraphics[width=\textwidth]{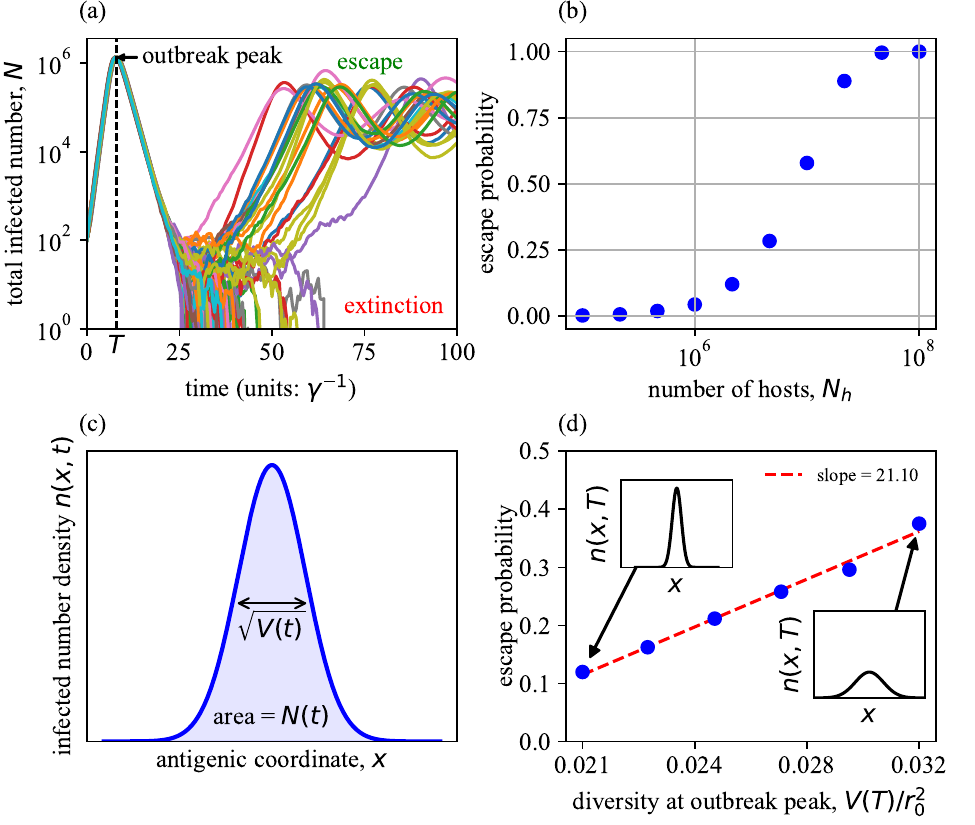}
    \caption{\textbf{For one deme, diverse pathogen populations better evade host immunity by accessing novel antigenic mutations.} \textbf{(a)} Total number of  infected hosts $N(t)=\int  n(x,t) dx$ for 200 replicate realizations of the model dynamics with an isogenic initial antigenic distribution. All replicates follow a similar trajectory until the accumulation of population size noise leads to one of two outcomes: pathogen escape or extinction. The timing of the peak of the outbreak is denoted by $T$.  \textbf{(b)} Escape probability as a function of the number of susceptible hosts. When the number of hosts is large, the probability of escape approaches 1. \textbf{(c)} Cartoon of the  density of infected hosts in antigenic space $n(x,t)$ during the early outbreak. The area under the curve gives the total number of infected individuals, $N(t)$ plotted in (a), while the variance of the antigenic density, $V(t)$, gives a measure of the diversity of antigenic strains within the population. \textbf{(d)} Probability of antigenic escape as a function of antigenic diversity at the time of the outbreak peak. We generate trajectories with different levels of antigenic diversity by starting with a Gaussian infected density, with different initial variances. We measure the diversity at the peak of the outbreak, $V(T)$  (time $T$ in panel (a)). Larger initial variance leads to an increased diversity at the outbreak peak. We quantify this relationship with a linear fit, $p=p_0+m \times V(T)$ (dashed line) with slope $m$. Parameters: $\beta = 2.5, \alpha = 0, \gamma = 1, D = 0.01, \sigma = 2, M = 15, N_h = 2 \times 10^6, r_0 = 3$, analysis was performed over $10^{4}$ statistical replicates for (b) and $3\times10^{3}$ replicates were used for (d). The initial variances were linearly spaced between $10^{-2}$ and $10^{-1}$ with 6 initial variances being used in total. 
    }
    \label{fig:one_deme}
\end{figure*}

To build intuition, we first study the simplest model that includes a single deme, eliminating spatial structure entirely. 
To study the effect of antigenic mutations on the outbreak dynamics, we numerically integrated Eqs. \eqref{viral_diff_eqn_please_work} and \eqref{immune diff eqn}.
Our initial antigenic distribution is isogenic with $N_0=100$ infected individuals all carrying the ancestor strain $x=0$, $n(x,t=0)=N_0\delta(x)$.
We find that the survival of the viral population is highly sensitive to noise stemming from population number fluctuations (Fig.~\ref{fig:one_deme} (a)).
The total number of infected individuals, 
\begin{equation}
N_1(t) = \int n_1(x,t)\,dx,
\end{equation}
follows two possible scenarios during an initial outbreak. After quickly rising to an outbreak peak, the infected host count either falls to zero as hosts gain adequate immune protection, leading to extinction (stochastic fade out) \cite{may1979population, anderson1979population,anderson1982directly, anderson1991infectious}, or rebounds if the viral population antigenically drifts sufficiently far from the ancestor strain to escape host immunity (Fig.~\ref{fig:one_deme}(a)). 

The escape probability depends on the total number of hosts, as we see by increasing the number of hosts in a one deme case (Fig.~\ref{fig:one_deme}~(b)). 
As $N_h \to \infty$, the viral survival probability in one deme approaches 1 because the virus has more time to develop crucial escape mutations before exhausting its supply of susceptible hosts.

Survival of the viral population also depends on the antigenic diversity of an outbreak, depicted schematically in Fig.~\ref{fig:one_deme}(c). The variance of the density of infected hosts $n_1(x,t)$, $V_1(t)=\av{x^2}_1-\av{x}_1^2$, where averages are taken over variants in deme 1 according to measure $n_1(x,t)/N_1(t)$,
 gives an estimate of the diversity of variants in the population. 
 Outbreaks with greater diversity have better access to novel mutations and are more likely to escape. 
 Seeding outbreaks with Gaussian distributed initial $n(x,t=0)$ with different values of $V(t=0)$, we record the diversity at time $T$ of the outbreak peak, $V(T)$. We find a linear dependence between the escape probability and the diversity at the outbreak peak (Fig.~\ref{fig:one_deme}(d)), showing that antigenic diversity influences viral survival.

\subsection{Two demes}

To see how the coupling of mobility and antigenic diversity influences viral escape, we turn to a two-deme system, with a constant migration rate $k$ between the two demes.
The off-diagonal elements of the migration matrix $k\equiv K_{12}=K_{21}$ in Eq.~\eqref{viral_diff_eqn_please_work} set the time scale over which the dynamics in the two demes equilibrate. 
We initialize our simulations with $N_0 = 100$ infected individuals in deme 1 as in the one-deme case, and none in deme~2.

\begin{figure*}
    \centering
    \includegraphics[width=\textwidth]{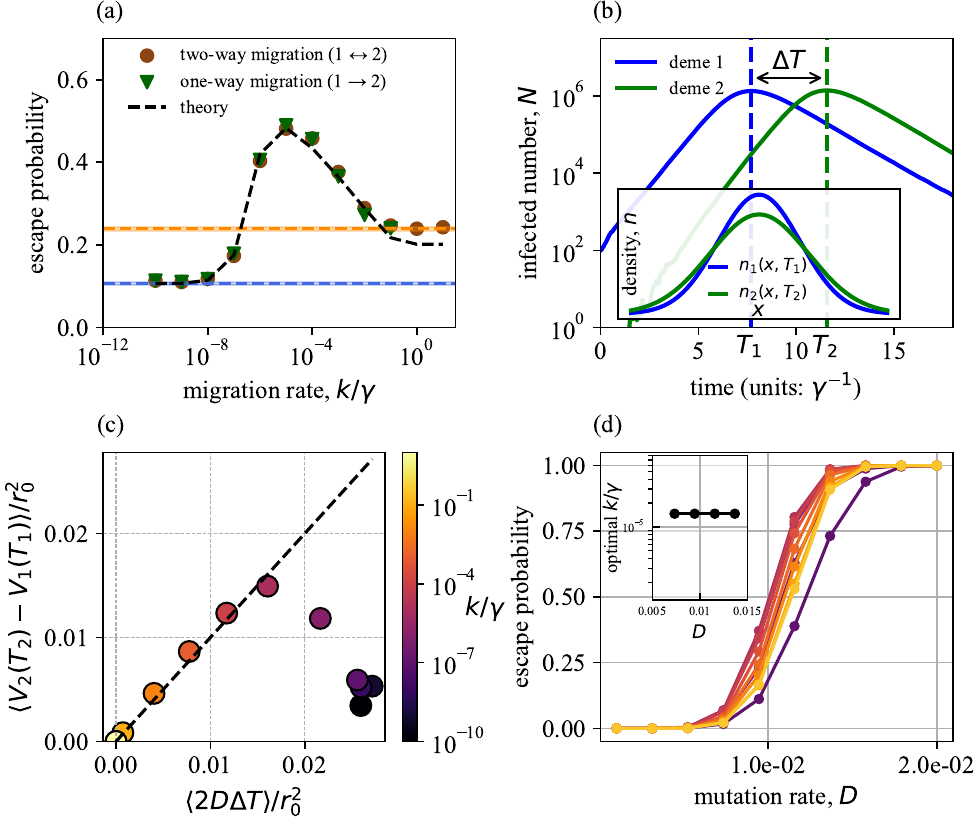}
    \caption{\textbf{Moderate host migration between two demes maximizes the probability of survival by amplifying the antigenic diversity of secondary outbreaks.} \textbf{(a)} The probability of antigenic escape at different migration rates for symmetric $1 \leftrightarrow 2$ (circles) and asymmetric $1 \rightarrow 2$ (triangles) migrations. The dashed black line is a prediction for the survival probability using the linear relation in Fig.~\ref{fig:one_deme}(d) and Eq.~\eqref{eqn:p_escape}. The horizontal lines show the result for the two limiting cases of zero migration rate (blue) and infinite migration rate (orange). The probability of survival at intermediate mobility attains a maximum near a rate of $ k/\gamma \approx 10^{-4}$. We do not report simulations of asymmetric migration for large $k$ because we neglect the time-dependence of $N_h$ which should only be important for large $k$. \textbf{(b)} Number of infected hosts in each deme for symmetric migrations. The outbreak is seeded in deme 1 with $100$ individuals, and spreads through host migration to deme 2 at a time $\Delta T$ in the future. The inset gives a comparison of the viral density of each outbreak at their respective peak, $n_1(x,T_1)$ and $n_2(x,T_2)$. The outbreak in deme 2 has large variance (i.e., is more antigenically diverse). \textbf{(c)} The increased  mean variance of the deme 2 outbreak  compared to deme 1 at their respective peaks, $\langle V_2(T_2)-V_1(T_1) \rangle$ as a function of the average time difference $\langle \Delta T \rangle$ between the outbreak peaks. Both axes have been scaled by the cross-reactivity $r_0^2$. The dashed line is the prediction $\Delta V = 2 D \Delta T$ (Eq.~\ref{eqn:additional_variance}). For small  migration rate, the time to seed an outbreak in deme 2, $\Delta T$, is larger than for large migration rate increasing the antigenic diversity, which increases the probability of antigenic escape as shown in Fig.~\ref{fig:one_deme}(d).  \textbf{(d)} Escape probability at different mutation and migration rates (denoted by the same color bar as in (c)). The inset shows that the optimal migration rate is independent of the mutation rate even though the overall escape probability depends strongly on $D$. Only mutation rates with $10^{-3} < \text{escape probability} < 1$ were used to compute the optimal migration rate. Parameters: $\beta = 2.5, \, \alpha = 0, \, \gamma = 1, \, D = 0.01, \, \sigma = 2, \, M = 15, \, N_h = 2 \times 10^6, \, r_0 = 3$ and the migration rate in panel (b) is $k = 10^{-3}$. In panels (a), (c), and (d) analysis was performed over $10^4$ statistical replicates.}
    \label{fig:two_deme}
\end{figure*}

At zero migration rate, individuals are unable to move between demes and the epidemiological dynamics in each deme are decoupled, replicating the case of a single deme discussed above. For infinite migration rates, spatial variation in the number of infected hosts is removed by rapid host movement and the combined population behaves as a single well-mixed population with twice the number of susceptible hosts. By calculating the proportion of trajectories in which the virus successfully escapes as a function of the spatial migration rate, we recover the expected behavior for small and large migration rates (Fig.~\ref{fig:two_deme}(a)). We also find that an intermediate migration rate maximizes the probability that the virus escapes host immunity, showing there is a preferred migration ratio that facilitates viral survival. The viral survival probability shows exactly the same dependence on the migration rate when we artifically forbid back migration of infected individuals from deme 2 to 1, $K_{21}=k$ and $K_{12}=0$ (Fig.~\ref{fig:two_deme}(a), triangles), suggesting that the success of the secondary epidemics in deme 2 is key to the virus survival.

To better understand, quantitatively, the interplay of migration and antigenic escape, we consider the timescales of outbreak peaks in the two demes for the simpler case of symmetric migrations. Fig.~\ref{fig:two_deme}(b) shows that the total number of infected hosts in each deme shows a time delay $\Delta T=T_2-T_1$ between their outbreak peaks. The distribution of infected hosts $n(x,t)$ in deme 2 at the outbreak peak is broader than in deme 1, signifying a higher diversity outbreak (inset of Fig.~\ref{fig:two_deme}(b)). The probability of antigenic escape for each migration rate is given by the probability of escape in each deme, $p_1$ and $p_2$, 
\begin{equation}
    p_{\rm escape} = 1 - (1 - p_1) (1 - p_2). 
    \label{eqn:p_escape}
\end{equation}
In the short time or low migration limit, we estimate the escape probability in deme 1, $p_1$, directly from one-deme simulations. The survival probability in deme 2, $p_2$, is the probability that the pathogen  escapes due to its increased diversity, given that it spread from deme 1 to deme 2. Due to the founder effect that gives an advantage to the very first migrating strains, the linear dependence of the escape probability on the antigenic diversity at the outbreak peak $V(T)$, as described in Fig.~\ref{fig:one_deme}(d), should also hold for deme 2, so that we may write
\begin{equation}\label{eq:p2}
    p_2=[p_1+m (V_2(T_2)-V_1(T_1))]p_{\rm spread},
\end{equation}
where $m$ is the regression coefficient, and $p_{\rm spread}$ is the probability that the epidemics spreads from 1 to 2.

For each value of the migration rate, we record the average antigenic diversity at the outbreak peak in the two demes, $\av{V_1(T_1)}$ and $\av{V_2(T_2) }$, and the frequency with which outbreaks in deme 2 are seeded via migration $p_{\rm spread}$.
The expected overall escape probabilities calculated using Eqn.~\eqref{eqn:p_escape} correctly predict the  measured probabilities for all but very high mutation rates (black line in Fig.~\ref{fig:two_deme}(a)) showing that the increased survival probability originates from an increased antigenic diversity in deme 2.

To understand why diversity in deme 2 is higher at its epidemic peak than in deme 1, we analyzed how the variances $V_1(t)$ and $V_2(t)$ are predicted to evolve according to Eq.~\ref{viral_diff_eqn_please_work}. 
During the first epidemic peak, immune pressure may be neglected and viral evolution in deme 1 is dominated by pure diffusion, so that $V_1(t)\approx 2Dt$. 
As soon as deme 2 receives some infected hosts from deme 1, its own diversity quickly tracks that of deme 1 when the migration is high enough, $V_2(t)\approx V_1(t)$, so that:
\begin{equation}
    V_2(T_2) - V_1(T_1) \approx 2 D \Delta T.
    \label{eqn:additional_variance}
\end{equation}
This prediction shows good agreement for small values of $\Delta T$, corresponding to large values of the migration rate $k$ (Fig.~\ref{fig:two_deme}(c)). Decreasing  the migration rate increases the time delay $\Delta T$, increasing antigenic diversity and the probability of survival of the virus in deme 2. As the migration rate becomes very low, and $\Delta T$ large, the secondary epidemics in deme 2 is seeded by only a few migrating hosts, leading to a collapse in the diversity at its peak. The effect may be accounted for qualitatively using simplifying approximations (see Appendix), yielding analytic or semi-analytic predictions for the diversity at the epidemic peak (solid lines in Fig. \ref{fig:two_deme}(c)). Combining these predictions with Eq.~\ref{eq:p2} allows us to derive an expression for the optimal migration rate, which decreases as a function of the magnitude of the epidemic peak. 
The feedback between migration rate and escape probability is entirely mediated by the outbreak delay time $\Delta T$, which should be independent of the mutation rate $D$.
In Fig.~\ref{fig:two_deme}~(d) we perform sweeps over both the mutation and migration rate and confirm that the optimal migration rate is insensitive to the choice of mutation rate. Outside of a small range around $10^{-2}$, the escape probability is either very close to one (high D) or very close to 0 (low D), making it difficult to see the effect of migration in our simulations. Within the range of interest, migration rate controls the sharpness of the transition, with the fastest rise occurring for the optimal migration rate. 

Altogether, the analysis of the two-deme case shows that the optimal migration should be slow enough to produce a delay in the outbreak dynamics, but fast enough so that the diversity of the initial epidemic in deme 1 can be inherited by deme 2. This effect entirely relies on the interplay of the migration and mutation rates.

\subsection{Linear networks with many demes}

\begin{figure*}
    \centering
    \includegraphics[width=\textwidth]{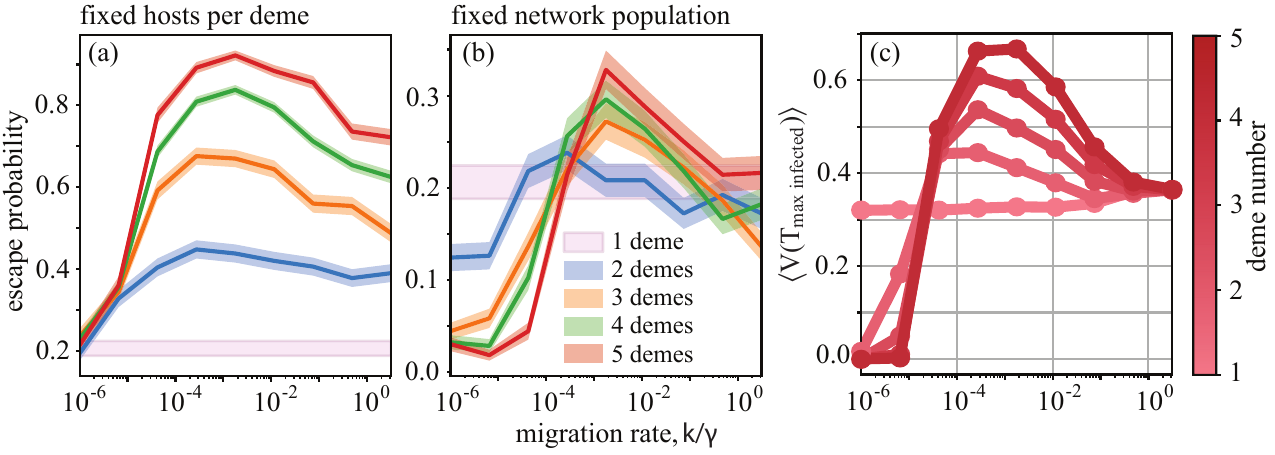}
    \caption{{\bf Survival probability increases with the number of demes}. \textbf{(a)}-\textbf{(b)} Survival probability as a function of the migration rate for networks with $N_D$ demes connected in a linear topology, with (a) a fixed number $N_h=10^6$ of hosts per deme (b) a fixed total population $N_{\rm tot}=10^6$ and $N_h=N_{\rm tot}/N_D$ hosts per deme. In both cases, symmetric migration rates are normalized by the number of outgoing connections a deme has, such that the rate out of any deme sums to $k$. \textbf{(c)} The mean variance of the viral antigenic density within each deme when the number of infected people in the deme is maximized in a five deme system with line topology - the same system producing the red line in (a) was used.}
    \label{effect of Nh}
\end{figure*}

To investigate networks with more than two demes and the role of subsequent deme outbreaks, we first consider the simple topology of $N_D$ demes arranged with linear connectivity, each deme with $N_h$ hosts. We assume symmetric rates normalized by the number of outgoing connections a deme has, such that the rate out of any deme sums to $k$. 
We observe a similar trade-off as in the two-deme case (Fig. \ref{effect of Nh}(a)), with an optimal migration rate achieving the balance between access to more hosts, and the ability for additional demes to give extra chances of survival to the virus.  Increasing the number of demes increases the escape probability at all migration rates, since it simply increases the number of hosts. This is clear in the limit of large migration rate ($k\to\infty$), which reduces to a single large deme of increasing size $N_D\times N_h$.

To disentangle the impact of deme structure from that of the number of hosts, we repeated the analysis but with constant total population size $N_{\rm tot}$, and deme sizes that decrease accordingly with the number of demes as  $N_h=N_{\rm tot}/N_D$ (Fig. \ref{effect of Nh}).
Despite this normalization, which decreases the overall survival probability, adding demes still increases viral survival at intermediate migration values (Fig. \ref{effect of Nh}(b)). The optimal migration rate for viral survival shifts slightly towards larger values as the number of demes increases. This result is explained by the number of hosts per deme decreasing, which makes survival in individual demes harder, and rescue through migration more important. 
Note that the large migration limit coincides with the single-deme result with $N_{\rm tot}$ hosts, as expected.

The benefit of adding more demes even as the total population size is kept constant emphasizes the importance of inheritance of diversity between demes. Diversity builds faster within a deme when the viral population is pressured by the immune system. When a new deme is seeded with the virus, there is no established immune system, and so not much diversity is generated inside it. Instead, diversity is imported from the previous deme, where it is already large, as additional infected hosts arrive.

For fixed numbers of hosts per deme (Fig.~\ref{effect of Nh}(a)), the optimal migration rate for  viral survival is the same for all  network sizes. As we argue in the appendix, the location of the peak decreases with the maximal number of hosts infected during the initial outbreak, $N_{\rm max}$. That peak is a property of each newly infected deme. It does not depend on the overall number of demes, but only on the number of hosts in the infected deme. When we add demes of constant size $N_h$ to the system, $N_{\rm max}$ remains similar for each deme, and so does the optimal migration rate. By contrast, when we increase the number of demes while keeping the total population constant, we expect the migration rate to shift to higher values as $N_h$ and thus $N_{\rm max}$ decrease, as observed in Fig.~\ref{effect of Nh}(b).

In Fig.~\ref{effect of Nh}(c) the importance of viral diversity for survivability in a larger deme system is revealed. The variance of the viral density is measured for each deme in a 5-deme line topology system. The last deme achieves the largest variance in the system, demonstrating the role of adding more demes in increasing viral diversity and therefore survival. Diversity in each deme is maximized at the migration rate at which survival probability also reaches a peak. This optimum occurs when the variance inherited by each successive deme is the same and the total variance within the system is also maximal. Consequently, the total diversity is optimized at a migration rate which allows for the same inherited diversity in all demes -- leading to the best chance of survival.

Overall, we conclude that separating hosts into many demes helps  the viral population survive. 

\begin{figure*}
    \centering
    \includegraphics[width=\textwidth]{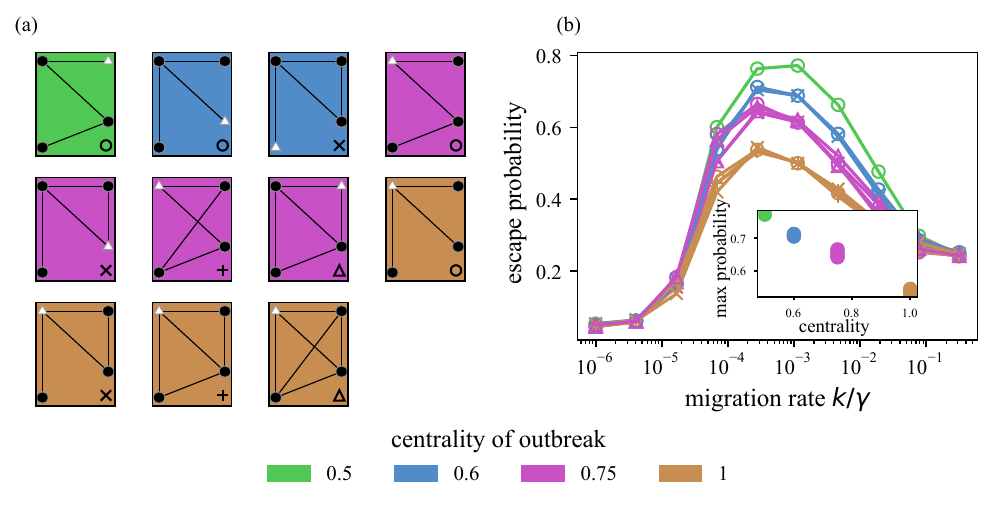}
    \caption{Effect of network topology on virus survival probability. \textbf{(a)} All four deme networks. White triangles indicate the deme/node with the initial outbreak. The color represents the value of the centrality of the outbreak node. Open symbols in the lower right corner of each colored box refer to the markers used in b), specifying the network topology used for the simulation of each curve.  Symmetric migration rates are normalized so that the total outgoing migration rate of each deme is always $k$. \textbf{(b)} Viral survival probability as a function of migration rate for all possible four-deme networks. Inset: peak height as a function of centrality. Simulation parameters: $\beta = 2.5$, $\alpha = 0$, $\gamma = 1$, $D = 0.01$, $r_0 = 3$, $\sigma = 2$, $M = 15$ $\sigma = 2$, $N_h = 10^6$.}
    \label{centrality networks}
\end{figure*}

\subsection{The influence of network topology on viral escape probability}
To explore the impact of topology beyond linear networks, we considered all unique four-deme network topologies (Fig.~ \ref{centrality networks}(a)), where uniqueness is defined with respect to network structure and the identity of the initial outbreak deme.  We assume symmetric migration rates normalized by the number of outgoing connections such that overall migration rate out of each deme is $k$. Fig.~\ref{centrality networks}~(b) shows the survival probability as a function of migration rate for these topologies. As in the previous cases, there exists an optimal migration rate at which the probability of survival is maximal, and the low and high migration rate limits reduce to the single deme case with $N_h$ and $N_{\rm tot}=4N_h$ hosts. The optimal migration rate for viral survival is relatively insensitive to network structure, since it is mostly driven by the deme size as discussed earlier. However, the value at the peak depends on the particular network topology.

To identify what features of the topology drive this dependence, we considered closeness centrality, a quantity that was proposed to quantify the importance of a node $x$ in a network, and defined as the inverse average distance to all other nodes in the system~\cite{Pastor-Satorras2015}: 
\begin{equation}
    C(x) = \frac{N-1}{\sum_{y=1}^{N-1}d(y,x)},
\end{equation}
where $N$ is the total number of nodes in the network and $d(y,x)$ is the length of the shortest path between nodes $x$ and $y$. 
We colored networks in Fig.~\ref{centrality networks}(b) and the survival probability curves in Fig.~\ref{centrality networks}(a) according to the closeness centrality of the outbreak node. The smallest closeness centrality value, represented by the line topology (in green), results in the highest  maximum survival probability. Networks of higher closeness centrality have lower survival probability peaks (inset of Fig.~\ref{centrality networks}~(a)), with the fully connected network having the lowest. Outbreaks with the same centrality have very similar curves.

Two aspects contribute to this effect: the network topology itself, and the location of the outbreak. Lower connectivities enhance virus survival, as do outbreak demes that are isolated.
Network  topology introduces an  exploration-exploitation tradeoff to viral survival. Outbreaks originating in nodes that are well connected gain access to more demes, and therefore susceptible hosts, early on. However, as the outbreak spreads quickly, only the diversity developed in the first deme is transmitted to other demes. Sequential discovery of new hosts in less connected networks allows the viral population to accumulate diversity via sequential range expansions. In the two deme case, access to new populations at a delayed time drives an increase in viral diversity. In larger networks, the structure of the network modulates the number of times the virus can exploit diversity accumulation due to this time delay. Outbreaks in well connected nodes expand to new populations simultaneously, while the linear structure of $N_D$ demes allows $N_D-1$ temporarily ordered range expansions that optimally exploit each deme.

\subsection{Where do real travel networks lie?} \label{real data}

\begin{figure*}
    \includegraphics[width=\textwidth]{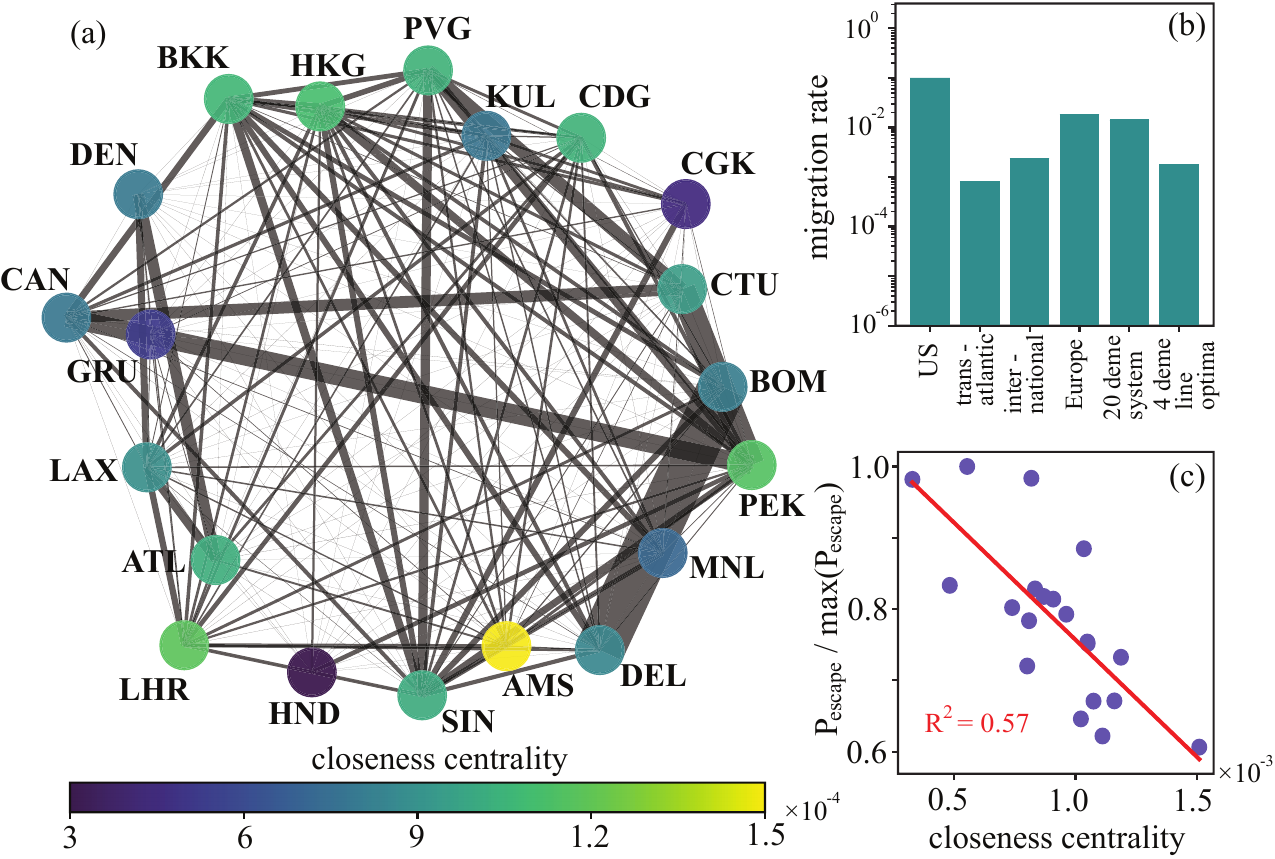}
    \caption{\textbf{Migration rates for real networks given flight travel data.} \textbf{(a)} Top 20 most frequently traveled airports from the data are displayed in a network graph with color gradient representing  the closeness centrality of each deme. \textbf{(b)} Migration rates are calculated for different real networks and are put into context with the four deme line topology migration rate which maximizes the survival probability. How these rates were obtained is outlined in the Supplemental Material. Data for the number of flights within designated regions is collected from ~\cite{statista2024}. \textbf{(c)} Inverse relationship between closeness centrality and probability of survival of the virus exists within the real network data. Parameters for simulation: $\beta = 2.5$, $\alpha = 0$, $\gamma = 1$, $D = 0.0042$, $r_0 = 3$, $\sigma = 2$, $M = 15$ $\sigma = 2$. Host population sizes in each deme reflect the metropolitan area population of the city where each airport resides ~\cite{macrotrends2024globalmetrics}.}
    \label{comp_real_net}
\end{figure*}

To determine if the relationship between closeness centrality and viral survival probability can be observed in real network data, we analyzed an airplane travel dataset~\cite{Worldpop2015}. The full dataset is composed of 3,632 airports and the number of people who travel between airport pairs in 2011. We select the top 20 most frequently traveled to airports and plot the network connectivity graph  with color coding according to the closeness centrality found for each node (Fig.~\ref{comp_real_net}(a)). The closeness centrality is calculated first by designating local distances between any two cities as the inverse of the number of people traveling between them. Using these distances as the directed weights in the network, Dijkstra's algorithm is used to find the shortest path between the origin deme and all other demes in the system. More explicitly, directed distance between demes $x$ and $y$ is the minimum over all paths between $x$ and $y$ of the sum of $k_{zz'}^{-1}$, where $zz'$ are edges of the path between $x$ and $y$. This places more strongly coupled airports closer together in network space. The network structure shows expected trends: the four Chinese airports (PEK, PVG, CAN, CTU) are strongly connected, as are  Mumbai (BOM) and Delhi (DEL) and the three US airports (ATL, LAX, DEN). We interpret each airport node of the graph as a deme. 

To understand how our model predictions connect to real world networks, we estimated travel rates based on the 20-deme network in Fig.~\ref{comp_real_net}(a), as well as other datasets ~\cite{statista2024}. Our estimates include only air travel, and are likely lower than the real values which include other modes of transportation.

We then compared these rates to the optimal migration rates obtained by the model for the four deme line topology, explored previously (Fig.~\ref{comp_real_net}(b)). While topologies of real networks are far bigger and more complex than our simulations allowed us to explore, our previous results suggest that the location of the maximum depends only weakly on the details of the network topology or size.
The comparison between the predicted optimal rate and the empirical estimates gives a good agreement, suggesting that current travel volumes maximally improve virus chances of survival.

To delve more finely into the network structure, we used the adjacency matrix produced by the twenty airport system outlined in Fig.~\ref{comp_real_net} ~(a) to  simulate and calculate the viral survival probability  for outbreaks initiated at each of the twenty airports. The migration rates used reflect the number of people moving between airports and are normalized by the metropolitan area population size of the origin deme~\cite{macrotrends2024globalmetrics}.  The viral escape probabilities are then plotted as a function of the closeness centrality of the airport where the outbreak originates.
The viral survival probabilities are negatively correlated with the closeness centrality (Fig.~\ref{comp_real_net}~(c)), meaning the more remote the outbreak deme, the greater the probability of survival. Therefore, we observe the same relationship in real world networks as in our smaller toy networks. The network graph and resulting ranking of demes by closeness centrality predict that an outbreak originating in the deme with the lowest closeness centrality, the only South American airport Saõ Paulo (GRU) in our reduced dataset, has the highest probability of survival. Conversely, a virus originating in Hong Kong (HKG), which is  well connected to other destinations around the world, benefits less from the ability to migrate. 

\section{Discussion}

When examining the interplay between viral evolution and host spatial migration we found that intermediate migration rates maximize viral diversity, favoring antigenic escape. This result stems from coupling a stochastic property of evolution - diversification -  and exploiting susceptible individuals due to migration. Large migration rates allow diversity to be transmitted between infected demes, as newly infected populations exert only weak or absent selection and multiple migration events allow multiple strains to simultaneously grow. This coupling leads to greater diversity and a higher probability of viral survival and ultimately antigenic escape. This process is an example of an exploration-exploitation tradeoff: low migration rates hamper viruses search for new pools of susceptible individuals, high migration rates result in infection with similar strains, decreasing diversity and future antigenic escape. This tradeoff depends on the ratio timescales for host immunity and migration. Low migration rates that result in exploiting a deme for long times allow the host immune systems to gain cohort immunity and eliminate viral diversity through selection. Long term survival of the virus therefore exploits short term coupling of host induced mutations and migrating at peak diversity. This process requires dwell times in a population that allow for the strain diversification but avoid low migration rates that would lead to a decrease in diversity before migration. 

The existence of an optimal migration rate for pathogen survival has been observed in a many models and attributed to different effects.  In our model, the optimal rate exploits the diversity build-up due to host-driven viral mutations, that are not included in models of long-term and large scale outbreaks~\cite{keeling2000metapopulation, pearce2020stabilization}. Matching allele models, originally introduced to study plant pathogen evolution, incorporate host/parasite migration, evolution, and demographics via a multicompartmental Lotka-Volterra style approach~\cite{gandon1996local}. However they assume fixed diversity. There are $n$ possible alleles for a single locus in both parasite and hosts, in which the host is immune to the ``matching'' parasite strain but susceptible to all others. Within this class of models, allowing only for host migration, Gandon~\cite{gandon1996local} reports an intermediate optimal value of host migration rates for host resistance. High pathogen persistence probabilities at intermediate pathogen migration rates have also been observed in two compartment Susceptible--Infected--Susceptible (SIS) models without an  evolutionary component~\cite{keeling2000metapopulation,Aleta_Hisi_Meloni_Poletto_Colizza_Moreno_2017,Hisi_Macau_Tizei_2019}.  Unlike these results, the optimal migration rate for viral survival in our model appears also in small networks, whereas in Hisi et al.~\cite{Hisi_Macau_Tizei_2019} the optimal migration rate disappears in networks of 10 or less demes.

The migration of susceptible hosts on the network has been discussed in terms of favoring one of two alternative rescue effects:  evolutionary~\cite{gandon1996local} or demographic~\cite{Hisi_Macau_Tizei_2019}. The evolutionary rescue effect relies on gene flow, where the genetics of a population is affected by the genetic makeup neighboring populations. This allows migrating strains that are poorly adapted to a given host environment to survive by migrating to new niches. It also allows well adapted populations to spread their genes to other regions, accelerating recovery after a shock.~\cite{lamarins2024eco}. This process is distinct from the demographic rescue effect, which creates short-lived refuges for viruses by randomly bringing a large number of susceptibles together for a period of time. The viral population can survive in this ephemeral host population until it reinfects other regions that have by then lost their immunity.  Our observations rely  on evolutionary rather than demographic rescue mechanisms since removing back-migration does not affect viral survival probabilities for a wide range of migration rates and viral survival correlates with large strain diversity. Overall, they reinforce the importance of evolutionary dynamics in metapopulation persistence, in line with \cite{mcmanus2021evolution,lamarins2024eco, thompson2019conflict}. 

A number of analytic studies have focused on ecological models without evolution, describing demographic rescue effects in two deme SIS systems. Meakin and Keeling~ \cite{Meakin_Keeling_2019} and Keeling and Rohani \cite{Keeling_Rohani_2002} use a moment closure approach to derive the correlation functions of the outbreak sizes in the demes.  These steady state results find highly correlated outbreaks- and therefore large maximum outbreak sizes- at intermediate coupling strength (both models use a different type of coupling: rather than permanently migrating, individuals ``commute" to the other deme for a fixed amount of time, then return home). Of course, survival probability and population size are not the same thing: our outbreaks peak after infecting most of the population, and then surviving simulations tend to recover and level off at about $10-20 \% N_h$ in each deme, roughly independent of migration rate. We also find that the outbreak peaks occur at the same time in different demes at high migration rates  (Fig.~\ref{fig:two_deme}~(c), $\Delta T$ goes to 0 at high migration rates), with a peak asynchrony (large $\Delta T$) at intermediate times. However, the composition of the outbreaks in each deme is not the same, which allows us to draw evolutionary conclusions from this observation. The outbreak in deme 2 is much more diverse than in deme 1 (see Fig.~\ref{fig:ridgeline_plot} in the appendix, \ref{Appendix_antigenic_diversity} for histograms of variance in by deme). Since diversity drives antigenic escape Fig.~\ref{fig:two_deme}~(b), comparison with this class of models also leads us to conclude that evolution is responsible for the optimal migration rate for viral survival.  At intermediate migration rates, the escaping strain appears in the more diverse deme 2, while at low migration rates is appears in deme 1, where the outbreak began (See Fig.~\ref{fig:which_deme_escapes} in the appendix, \ref{Appendix_antigenic_diversity} ).

Evolutionary models without migration have predicted a nonlinear increase in the speed of antigenic evolution as a function of the fraction of immunocompromised hosts  in the population ~\cite{kumata2022antigenic}. This result is due to hosts that recover more slowly infecting more people, as they have more time to transmit the infection. This causes an effective increase in the growth rate of the virus, increasing the probability of acquiring an antigenic mutation. Immunocompromised hosts play a similar role as migration in our model: when a new deme is infected for the first time the virus has access to  a large number of hosts with susceptible immune systems,  increasing the growth rate of the virus, but only in a transient way. While we do not systematically study the effect of base virulence $\beta$ in our model, previous theoretical work has shown that shorter, more virulent outbreaks tend to produce less genetic diversity than longer, less infectious ones~\cite{Boni_Gog_Andreasen_Feldman_2006}. That is, the virus needs time to exploit mutations and time is in short supply in the fast, sudden and one-time outbreaks that happen with high $\beta$. Comparison to this class of models allows us to draw conclusions about the evolutionary impact of population structure in space.

The topology of the network determines the probability of survival of the viral population.  Outbreaks in demes with low closeness centrality  are more likely to escape the host immune systems than ones starting in  high closeness centrality demes. This means that the way a virus disperses has evolutionary consequences, at least when evolution happens quickly. While we focused on small networks, these effects likely hold in large networks. If this is the case, not all initial conditions of the simulation are equal. In the case of randomly chosen initial conditions over many replicates, not all replicates contribute equally to observed mean probabilities. Depending on how extreme this effect is, some initial outbreak locations may dominate the viral survival probabilities on large networks. In the context of protecting crops from dispersing pests, it has been noted that the degree of the outbreak node is correlated with outbreak size and the speed of spread ~\cite{Pautasso_Moslonka-Lefebvre_Jeger_2010,Gent_Bhattacharyya_Ruiz_2019}. 

The role of topology for viral survival also suggests that multiple pairwise mobility restrictions between cities  or airports can synergistically strengthen and result in a stronger global effect. This has been reported when implementing spectral control strategies~\cite{Bishop_Shames_2011,darabi2023centrality}, which selectively modify traffic between specific nodes to control epidemics. In the long time limit, SIS models on networks have shown that the maximal eigenvalue of the network determines the exponent of the rate with which a particular virus goes extinct \cite{Pastor-Satorras2015}. Our model complements these results with analysis of the transient dynamics. The networks that favor antigenic escape in the short term are not necessarily the same as those that favor long term survival. For example, the linear network topology guarantees the largest probability of short term antigenic escape, but has a small maximal eigenvalue, which makes viral survival more difficult 
 at long times. The network that makes viral survival most difficult at short times, the fully connected network, has a large maximal eigenvalue and offers the best conditions for long term viral survival. Therefore, the traffic restriction strategies which are most effective will be different for emergent and endemic pathogens. 

In the case of emergent pathogens, the applicability of our results will depend on the scale considered. Fig.~\ref{comp_real_net} shows that depending on which set of airport data one considers, the effective  migration rate is not the same. All reported migration rates result in viral survival probabilities that are close to the maximum (Fig.~\ref{comp_real_net}~(b)). This result could imply  that reducing viral survival could be achieved either by strongly increasing or decreasing mobility. However, the value of viral survival also depends on the number of  hosts (Fig.~\ref{effect of Nh}~(c)). If the number of hosts is large, increasing migration rates does not decrease the viral survival probability. Additionally, the width of the  peak in terms of migration rates means obtaining a significant effect would require changing mobility by several orders of magnitude, which is no small task for policymakers. Additionally, while decreasing mobility is an achievable policy goal, increasing mobility is harder to implement. For example, Refs.~\cite{Chang_Pierson_Koh_Gerardin_Redbird_Grusky_Leskovec_2021} report that cell phone mobility data shows that  the "lockdown" policy caused a 54 percent reduction in mobility in the Chicago metro area in the first week of April 2020. Lastly, driving a virus to extinction is not the primary goal of policy. It is more important to reduce the total number of cases and to prevent an outbreak large enough to overwhelm the healthcare system. 

Eco-evolutionary feedbacks remain critical to understanding long term epidemiology. We have examined a density-dependent selection effect (very common strains within a deme experience the most immune pressure, neglecting the lag time for collective immunity to catch up), but global frequency dependent effects can be important as well. It also remains to be seen how short time selection biases towards diversity impact mid and long term phylodynamics. Such a question requires speciation dynamics and therefore a higher dimensional antigenic space. 
 
\section*{Acknowledgements}
This work was carried out as a summer school project during the Les Houches summer school on Theoretical Biophysics. AMW and TM thank Alain Barrat for useful discussions. This work was supported by the European Research Council
consolidator grant no 724208 (AMW, TM), and the Agence Nationale
de la Recherche grant no ANR-19-CE45-0018 “RESP-REP” (AMW, TM). AI-R received funding from the European Union’s Horizon 2020 research and innovation programme under the Marie Skłodowska-Curie grant agreement No 847718. DS acknowledges the MIT Engaging cluster for providing computational resources and support. NB acknowledges funding from the EMERGENCE(S) grant program of the city of Paris. This publication reflects only the authors' views. The relevant funders are not responsible for any use that may be made of the information it contains.

\section{Appendix}
\subsection{Oligomorphic Dynamics} \label{sec:OMD}
We adapt the Oligomorphic Dynamics (OMD) approximation \cite{lion2023extending, sasaki2011oligomorphic} to study the size and antigenic variance of the initial outbreak. 
For simplicity we  neglect noise and work in two demes. 
The dynamics of the densities in deme $i$ are given by Eqs.~\eqref{viral_diff_eqn_please_work}, \eqref{immune diff eqn} \eqref{ceqn}, and \eqref{Seqn}, where we define the fitness as $F_i(x,t) = \beta S_i(x,t) - \alpha - \gamma$ and, for simplicity we  neglect noise and work in two demes. We assume that there are $N_0$ individuals infected at time $t=0$, all with strain $x=0$ in deme 1, which gives the initial condition $n_1(x,t=0) = N_0 \delta(x)$ and $n_2(x, t=0)=0$. We  obtain an equation of motion for the total number of infected individuals $N_i(t)=\int dx n_i(x,t)$ by integrating  Eq.~\eqref{viral_diff_eqn_please_work}:
\begin{equation}
    \frac{d N_i}{dt} = \langle F_i \rangle N_i + k(N_j - N_i),
    \label{EqnSI_dNdt}
\end{equation}
The total infected population grows at the antigen-averaged fitness and migration causes the infected numbers in different demes to equilibrate on a timescale of $k^{-1}$. 

We denote the densities of viral variants and immune protections as:
\begin{align}
    \phi_i(x,t) &= \frac{n_i(x,t)}{N_i(t)} \label{OMD2} \\
    \psi_i(x,t) &= \frac{h_i(x,t)}{H_i(t)},
    \label{OMD}
\end{align}
with $H_i(t)=\int dx\, h_i(x,t)$.

Applying Eqs.~\eqref{OMD2} and \eqref{OMD} to Eq.~\eqref{viral_diff_eqn_please_work} we obtain:
\begin{eqnarray}
    \frac{\partial \phi_i}{\partial t} &=& \frac{1}{N_i} \frac{\partial n_i}{\partial t} - \frac{n_i}{N_i^2}\frac{d N_i}{dt}\\ \nonumber 
    &&= (F_i - \langle F_i \rangle ) \phi_i + D \frac{\partial^2 \phi_i}{\partial x^2} + \frac{k N_j}{N_i}(\phi_j - \phi_i).
    \label{phieqn}
\end{eqnarray}
The first term tells us that regions of antigen space with fitness larger than average will increase in frequency while regions with fitness less than average will to decrease in frequency. The migration term has a  $N_j / N_i$ prefactor. 
If deme $j$ has a much larger infected population than deme $i$, the antigenic distribution in deme $i$ will rapidly relax to the antigenic distribution of deme $j$,  inheriting all the genetic diversity from deme $j$ almost immediately. This rapid equilibration of the antigenic distributions among demes explains the enriched diversity vs outbreak size relationships for secondary outbreaks we see in simulation. 

We now assume $\phi(x,t)$ is  Gaussian and can thus be characterized completely by its mean $m_i(t) = \langle x \rangle_i$ and variance $V_i(t) = \langle (x - \langle x \rangle_i )^2 \rangle_i$. 
We assume the distribution of $x$ is sharply peaked about the mean value $m_i(t) = \av{x}_i$ and expand the fitness around the mean:
\begin{eqnarray}
    \langle F_i(x) \rangle_i& \approx& \langle F_i(m_i) + F_i'(m_i) (x-m_i) \\ \nonumber 
    &&    + \frac{1}{2}F''(m_i)(x-m_i)^2 + ...\rangle_i \\ \nonumber 
    &&= F_i(m_i) + \frac{F_i''(m_i)}{2}V^2_i + ...,
\end{eqnarray}
where we neglect higher order terms than the variance $V_i(t) = \av{(x - \langle x \rangle_i )^2}_i$. This approximation is valid since for typical simulation parameters, $V/r_0^2 \sim 2Dt/r_0^2 \sim 10^{-2}$, which renders even the higher order terms small compared to first and second order terms. 
The OMD approximation for the total infected number gives:
\begin{equation}
    \frac{dN_i}{dt} = \left(F_i(m_i) + \frac{F_i''(m_i)}{2} V_i\right) N_i + k (N_j - N_i).
\end{equation}

To close the equation we  need dynamics for the mean and variance, $m_i$ and $V_i$. $ m_i = 0$ at all times based on symmetry arguments. Multiplying Eq.~\ref{phieqn}  by $x^2$ and integrating by parts we find 
\begin{align}
    \frac{d V_i}{dt} &= \int x^2 \left((F_i - \langle F_i \rangle ) \phi_i + D \frac{\partial^2 \phi_i}{\partial x^2} + \frac{k N_j}{N_i}(\phi_j - \phi_i)\right) dx \nonumber \\
    &= \text{Cov}(x^2,F_i(x)) + 2D + k\frac{N_j}{N_i}(V_j-V_i). \label{eqn:dvidt}
\end{align}
This is a generalization of the Price equation to include  diffusive spreading due to mutations and the rapid distribution equilibration due to migration.
 
\subsection{Deriving $\Delta V = 2 D \Delta T$}

For the case of two demes, the covariance term in Eq.~\ref{eqn:dvidt} scales with $V_i^2$ at early times when $V_i$ is still small, and can thus be neglected,
giving equations for the variance in the OMD approximation
\begin{align}
    \frac{dV_1}{dt} &= 2D + k \frac{N_2}{N_1}(V_2-V_1)\label{eq:div2} \\ 
    \frac{dV_2}{dt} &= 2D + k \frac{N_1}{N_2}(V_1-V_2) . 
\end{align}
If the outbreak starts in deme 1 and spreads to deme 2 at $\Delta T$, at small times we have $V_1 \approx 2 D t$ and the solution of
\begin{align}
    \frac{dV_2}{dt} &= 2D + k \frac{N_1}{N_2}(2Dt - V_2) \\
    V_2(\Delta T) &= 0
\end{align}
is
\begin{equation}
    V_2(t) = 2D \left(t - \Delta T e^{-k\int_{\Delta T}^{t} \frac{N_1(t)}{N_2(t)} dt}\right).
\end{equation}

If the infected number of individuals in deme 1, $N_1$, peaks at time $T_1$, and the infected number in deme 2 peaks at time $T_2 = T_1 + \Delta T$, 
the antigenic variance at the outbreak peak in deme 2 is 
\begin{equation} \label{sieqn:vardiff}
    V_2(T_1 + \Delta T) = 2D T_1 + 2 D \Delta T \left(1 - e^{-k\int_{\Delta T}^{T_1 + \Delta T} \frac{N_1(t)}{N_2(t)} dt}\right).
\end{equation}
For large $k$, the spread to deme 2 happens quickly, $\Delta T \approx 0$, such that
\begin{equation}
    V_2(T_2) \approx 2 D (T_1 + \Delta T) = V_1(T_1) + 2 D \Delta T.
\end{equation}

\subsection{Antigenic Diversity in each deme}
\label{Appendix_antigenic_diversity}
We have argued in the main text that antigenic diversity drives our results. Here we include more detailed simulation results from the 2 deme case to demonstrate that diversity is systematically higher in the second deme. We compare diversity between demes at comparable points in the development of the outbreak in each deme, i.e. the diversity of each deme at the peak of the outbreak in each deme. 
This amounts to comparing the within deme diversity with the time lag of $\Delta T$. 

\begin{figure}
    \centering
    \includegraphics[width=\linewidth]{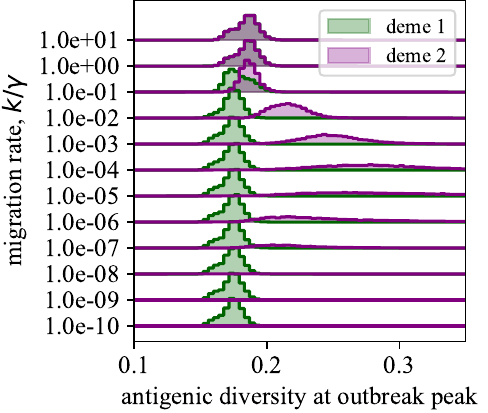}
    \caption{Histograms of diversity at outbreak peak in each deme show that at optimal migration rates, deme 2 is markedly more diverse than deme 1. The histograms are count normalized, so low migration rates which rarely seed secondary outbreaks produce very few samples of diversity in deme 2.}
    \label{fig:ridgeline_plot}
\end{figure}

Fig.~\ref{fig:ridgeline_plot} shows the histogram of antigenic diversity in each deme at outbreak for a sweep of migration rate $k$. 
At low migration rates (bottom) there are very few outbreaks in deme 2, so the deme 2 histogram is very sparse. As migration rate increases and becomes close to the optimal migration rate for viral survival, we see that deme 2 tends to have more diverse outbreaks, with the mean of the distribution shifting right as well as its width growing to close to four-fold compared to deme 1. Deme 2 is much more likely to see very diverse outbreaks than deme 1. Increasing $k$ further, rapid mixing between the demes takes over and the distributions become more similar. Here invasion to too rapid to exploit deme 1 to gain diversity and the virus immediately explores both demes. 

\begin{figure}
    \centering
    \includegraphics[width=\linewidth]{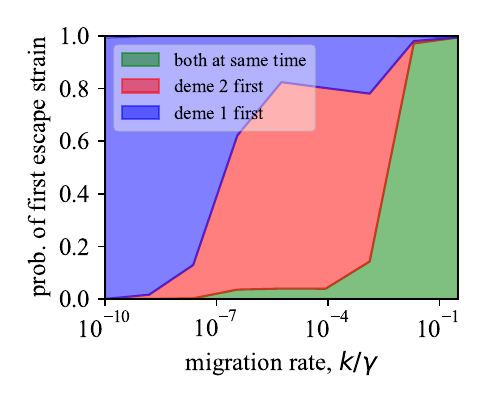}
    \caption{Probability of escape occurring in each deme for various migration rates in the two deme case. The size of the red region gives the probability that the antigenically escaped strain first appears in deme 2 as opposed to deme 1. We define an antigenically escaped strain as one which is sufficiently far from the ancestor strain at $x=0$. In deme $i$, the time of emergence is the first time $t$ such that $\int_{|x|>x^*} n_i(x,t) dx > 100$ with $x^*$ (the critical antigenic distance) computed as in Ref.~\cite{sasaki2022antigenic}. If the new strain emerges in both demes within 2 time units of one another, we say they appear at the same time. We highlight that at intermediate migration rates deme 2 is the primary source of escape strains due to its amplified antigenic diversity. Parameters are the same as in Fig.~\ref{fig:two_deme}}
    \label{fig:which_deme_escapes}
\end{figure}

The correlation of higher diversity with antigenic escape is demonstrated by recording which deme produces an escape variant first. We use OMD arguments to set an escape distance, $x*$, such that immunity profile created in response to an outbreak centered at 0 has largely decayed. We define the escape time $T_e^i$ for each deme $i$ as the first time a virus appears in that region of antigenic space. We then compare the escape times in each deme to see, over many runs, where the escape is most likely to occur (Fig.~\ref{fig:which_deme_escapes}). At low migration rates, deme 1 has escape events more often than deme 2, as deme 2 is more rarely populated. At intermediate migration rates, the more diverse deme 2 has escape events  more often than deme 1, and at high migration rates the two demes show similar numbers of escape events. 

\subsection{Numerical procedures}
Our simulations produce solutions to the system of stochastic integro-PDEs defined by Eqs.~\eqref{viral_diff_eqn_please_work}, \eqref{immune diff eqn}, \eqref{ceqn}, and \eqref{Seqn}, with all codes to produce simulation results and figures available on GitHub.

Our simulation method draws on previous work, mainly references \cite{pechenik1999interfacial, weissmann2018simulation, dornic2005integration}, which we will review here. 
The key idea behind our numerical solver is the ``operator splitting method" \cite{issa1986solution, macnamara2016operator}, where the terms in \EQ{viral_diff_eqn_please_work} are solved one at a time - sequentially for each time step. 
We split \EQ{viral_diff_eqn_please_work} into a deterministic and a stochastic piece, and solve each individually. 
We will solve our system on a uniform grid of antigenic points with grid spacing $\Delta x$ and periodic boundary conditions. 
The length of the antigenic space is $L$, so the antigenicity can take values $x = -L/2, -L/2 + \Delta x, ... , L/2 - \Delta x$.
We assume periodic boundary conditions so that antigenic site $x=L/2$ is identical to $x = -L/2$, but other choices of boundary condition (such as no flux) are possible. 
We choose values of $L$ large enough so that the infected density never interacts with the boundary of the antigenic space. 
As in the continuum model, we will call the infected density in deme $i$ at antigenic coordinate $x$ and time $t$, $n_i(x,t)$, the immune density $h_i(x,t)$, the cross-reactive immune protection $c_i(x,t)$ and the susceptibility $S_i(x,t)$.

Time is likewise discretized by steps of length $\Delta t$, for the deterministic step of the algorithm we use the explicit forward-time Euler method to update all densities. 
In a given time step $\Delta t$, we first perform a diffusion update to capture mutational changes. At each antigenic point $x$ we compute
\begin{multline}
    \Tilde{n}_i(x,t) = n_i(x,t) + 
    \frac{D \Delta t}{\Delta x^2} \Big( n_i(x+\Delta x, t) \\
    + n_i(x - \Delta x, t) - 2 n_i(x,t) \Big).
\end{multline}
We then compute the cross-reactive immune protection $c_i(x,t)$ at each grid point $x$
\begin{equation}
    c_i(x,t) = \sum_y \exp\left(\frac{\text{min}(|x-y|,L-|x-y|)}{r_0}\right) h_i(y,t) \Delta x,
\end{equation}
where the sum runs over $y = -L/2, -L/2 + \Delta x, ... , L/2 - \Delta x$ and the minimum function accounts for the periodic boundary conditions on antigenic space.
The cross-reactive protection $c_i(x,t)$ is then used to compute the susceptibility $S_i(x,t) = (1 - c_i(x,t))^M$.
We next update the immune and infected densities using the post-mutation infected density $\Tilde{n}_i(x,t)$,
\begin{align}
    h_i(x,t+\Delta t) &= h_i(x,t) + 
    \frac{\Delta t}{M N_h} \Big( \Tilde{n}_i(x,t) \nonumber \\
    &\quad - \Tilde{N}_i(t) h_i(x,t) \Big) \\
   \ttilde{n}_i(x,t) &= \Tilde{n}_i(x,t) \left( 1 + \Delta t F_i(x,t) \right),
\end{align}
where 
\begin{equation}
    \Tilde{N}_i(x,t) = \sum_x \Tilde{n}_i(x,t) \Delta x
\end{equation}
and
\begin{equation}
    F_i(x,t) = \beta S_i(x,t) - \alpha - \gamma.
\end{equation}

We now turn to treat the stochastic piece of the dynamics.
The characteristic feature of demographic noise are fluctuations that scale like the square root of the population size. 
Between any two grid points in antigenic space $x \to x + \Delta x$, we expect there to be $n_i(x,t) \Delta x$ infected individuals in deme $i$.
We then expect that the noise amplitude should be proportional to $\sqrt{n_i(x,t) \Delta x}$.
The factor of $\Delta x$ here is important, previous work \cite{pechenik1999interfacial, weissmann2018simulation, dornic2005integration} assumes a unit lattice spacing which is appropriate for a stochastic PDE. 
Our model by contrast requires direct integration over the antigenic coordinate in order to compute immune response. 
This integration requires definition of an antigenic step size $\Delta x$ which we eventually make small.
We begin by writing the stochastic differential equation (SDE) for the number of infected individuals between $x$ and $x+\Delta x$
\begin{equation}
    \frac{d}{dt}(n_i(x,t) \Delta x) = \sigma \sqrt{n_i(x,t) \Delta x} \xi_i(x,t).
\end{equation}
The process $\xi_i(x,t)$ is a white noise with correlation function
\begin{equation}
    \langle \xi_x(x,t) \xi_j(y,s) \rangle = \delta_{ij} \delta_{xy} \delta(t-s)
\end{equation}
The SDE for the density at grid point $x$ is then
\begin{equation}
    \frac{d n_i}{dt} = \frac{\sigma}{\sqrt{\Delta x}} \sqrt{n_i} \xi_i(x,t).
\end{equation}
The solution to this SDE is known \cite{dornic2005integration}, we directly integrate from $t$ to $t+\Delta t$ where the initial condition is the result of the deterministic step $\tilde{n}_i(x,t)$. 
The distribution for $n_i(x,t+\Delta t)$ is
\begin{align} \label{eq:PLIntegrator}
    n_i(x, t+\Delta t) &\sim \text{Gamma} \Bigg( 
    \text{Poisson}\left( \frac{2 \Tilde{n}_i(x,t) \Delta x}{\sigma^2 \Delta t} \right) \Bigg) \nonumber \\
    &\quad \times \frac{\sigma^2 \Delta t}{2 \Delta x}.
\end{align}
The argument of the Gamma distribution above is the scale parameter, the rate parameter is 1.
Other methods of sampling the noise are also possible \cite{chardes2023evolutionary} and produce qualitatively identical behavior, with a common choice being a simple Poisson sample
\begin{equation} \label{eq:PoissonIntegrator}
    n_i(x, t+\Delta t) \sim  
    \text{Poisson}\left( \frac{ \Tilde{n}_i(x,t) \Delta x}{\sigma^2 \Delta t} \right) \times \frac{\sigma^2 \Delta t}{ \Delta x}.
\end{equation}
In the main text, all simulations except for those of Fig.~\ref{effect of Nh} were performed using Eq.~\eqref{eq:PLIntegrator}.
Data for Fig.~\ref{effect of Nh} were generated using Eq.~\eqref{eq:PoissonIntegrator}
The final step is to include spatial migrations, which we incorporate using a mass-action term
\begin{align}
    &n_i(x,t+\Delta t) \leftarrow n_i(x,t+\Delta t) \nonumber \\
    &\quad + \sum_j \left[ K_{ij} n_j(x,t+\Delta t) - K_{ji} n_i(x,t+\Delta t)\right]\Delta t.
\end{align}

\subsection{Code availability}

All codes used to perform simulations and generate figures can be found on GitHub: \url{https://github.com/caelan-brooks/viral-coev-net}.

\subsection{Computing the optimal migration rate} \label{section:compute_delta_T}

The first step to determining the optimal migration rate is to calculate how long on average it takes for an outbreak in deme 1 to spread via migration to deme 2. 
The model defined by Eqs.~\eqref{viral_diff_eqn_please_work}, \eqref{immune diff eqn}, \eqref{ceqn}, and \eqref{Seqn} is complex, we will make simplifying assumptions when necessary in order to make progress. 
We will use the OMD framework of \ref{sec:OMD} to compute the number of infected individuals and antigenic diversity ($N_i, V_i$) in each deme.
We will assume that in deme 1 the viral abundance follows an exponential growth followed by an exponential decay 
\begin{equation}
    N_1(t) =
\begin{cases} 
N_0 e^{F t} & \text{if } t \leq T_1 \\
N_{max} e^{-\gamma t} & \text{if } t > T_1
\end{cases}.
\end{equation}
This says that the infected number in deme 1 will grow exponentially with rate $F = \beta - \gamma$ until reaching a maximum at the outbreak peak time $T_1$.
At this time, there are $N_{max} = N_0 e^{FT_1}$ infected individuals and the infected population will then decay as individuals recover at rate $\gamma$. 
In deme 2, migrant infected populations will be subject to growth and noise.
Our primary interest is in the emergence of the infected population in deme 2 as it grows to avoid stochastic extinction. 
We will assume that the small viral populations in deme 2 during the early outbreak elicit negligible immune response, instead enjoying a constant growth rate $F$. 
Integrating Eq.~\eqref{viral_diff_eqn_please_work} over $x$ and ignoring migrations, we find 
\begin{equation} \label{sieqn:dN2dt}
    \frac{dN_2(t)}{dt} = F N_2 + \sigma \sqrt{N_2} \xi(t),
\end{equation}
where $\xi(t)$ is a white noise.
Using Eq.~\eqref{sieqn:dN2dt}, we can compute the probability that a virus with initial abundance $N_2(0)$ eventually goes extinct due to noise. 
The calculation is fastest using common results from martingale theory \cite{roldan2023martingales}.
The stochastic process $Z(t) = \exp(-2 F N_2(t) / \sigma^2)$ is a martingale which is shown by computing $dZ/dt$ using It\^o's lemma and noticing that $Z(t)$ has no deterministic rate of change. 
The process $Z(t)$ being a martingale implies that its expectation should not change in time (with expectation $\langle ... \rangle$ here being over noise realizations, not over the antigenic distributions as previously discussed).
Because $Z(t)$ has a known deterministic initial condition, we immediately have 
\begin{equation}
    \begin{split}
        \langle Z(t) \rangle = Z(0) &= P(\text{extinction}) \exp\left(-\frac{2F \times 0}{\sigma^2}\right) \\
        &\quad + P(\text{establishment}) \exp\left(-\frac{2F \times \infty}{\sigma^2}\right)
    \end{split},
\end{equation}
which immediately gives the probability of establishment ($1-P(\text{extinction})$) is 
\begin{equation}
    P(\text{establishment}) = 1 - Z(0) \approx \frac{2 F}{\sigma^2} N_2(0) \label{sieqn:p_est}.
\end{equation}
To include migration, we note that in a given time interval $t \to t + dt$, $k \times N_1(t)$ individuals migrate from deme 1 to deme 2. 
Calling the time at which the first successful migration arrives in deme 2 $\Delta T$, we can explicitly calculate
\begin{align}
    P(\Delta T > t) &= \prod_{s<t} \left(1 - \frac{2F}{\sigma^2} k N_1(t)\, dt \right) \\
    &= \exp\left( -\int_0^t \frac{2kF}{\sigma^2} N_1(s) \, ds \right) \label{sieqn:cdf}.
\end{align}
Computing the mean of a random variable distributed according to \eqref{sieqn:cdf} is difficult in general. 
Instead, we compute the most likely value for $\Delta T$, which is the inflection point for the cdf, $\frac{d^2P(\Delta T > t)}{dt^2} = 0$.
The most likely value for the migration time $\Delta T$ is given by
\begin{equation}
    N_1(\Delta T)^2 = \frac{\sigma^2}{2kF}N_1'(\Delta T).
\end{equation}
Assuming that migration happens before the outbreak peaks in deme 1 ($\Delta T < T_1$) we fined a simple expression for the migration time
\begin{equation}
    \Delta T = \frac{1}{F}\log\left( \frac{\sigma^2}{2kN_0}\right). \label{sieqn:Tvsk}
\end{equation}
We now have a concrete relationship between the rate of host migration and the time to seed an outbreak in deme 2. 
One could also compute the expected value of $\Delta T$ numerically, $\langle \Delta T \rangle = \int_0^\infty P(\Delta T > t) dt$, which we implement when comparing to simulation results.
Once the first infected individual as successfully established in deme 2, the growth should be approximately exponential 
$N_2(t) = e^{F(t-\Delta T)}$. 

Taking $N_2$ to be identical to $N_1$ but shifted in time by an amount $\langle \Delta T \rangle$, we can evaluate the integrals in Eq.~\eqref{sieqn:vardiff}, we find
\begin{align}
    &V_2(T_1 + \langle \Delta T \rangle) - V_1(T_1) 
    \approx \nonumber \\
    &2D \langle \Delta T \rangle \left( 1 - \exp\left( - k N_0 e^{F \langle \Delta T \rangle}  \left( T_1 - \langle \Delta T \rangle + \beta^{-1} \right) \right) \right) \nonumber \\
    &\approx 2D \langle \Delta T \rangle \left( 1 - \exp\left( - \frac{\sigma^2}{2} \left( T_1 - \langle \Delta T \rangle + \beta^{-1} \right) \right) \right).
\end{align}
The above equation can be further simplified by applying Eq.~\eqref{sieqn:Tvsk}, and maximizing the resulting variance difference as a function of $k$.

\subsection{Migration rate predictions}

In Figure ~\ref{comp_real_net}b we compare migration rates between different regions in the world to see what probability of survival would have in these systems. To do this, we obtain data on the number of flights per year in within that system (for the US calculation we would find the number of internal US flights per year) as well as the total number of people in that system (number of people living in the US) and the average number of passengers on a plane flying within the region. The calculation would follow: 

\begin{align*}
    &\text{Migration rate} = \\
    &\frac{(\text{\# of flights per year})(\text{average \# of people per flight})}
    {(\text{52 weeks per year})(\text{\# of people in region})}
\end{align*}

We end up with the number of people moving per week / total people in the region as a migration rate prediction. The number of people traveling per week is used due to the migration rate scaling with the recovery time which is about a week. 

\subsection{US migration rate calculation}

\begin{align*}
    & \text{US migration rate} = \\
    & \frac{(16,405,000  \text{ passengers per year})(200 \text{ passenger per flight})}{(52 \text{ weeks per year})(333,300,000 \text{ US residents})} \\
    & \hspace{3cm} = 2 \text{ x } 10^{-1}
\end{align*}

From this calculation, travel within the US is well mixed.

\bibliographystyle{pnas}

\end{document}